# Gigavoxel-Scale Multiple-Scattering-Aware Lensless Holotomography

Mikołaj Rogalski[1,~], Julianna Winnik[1,~], Julia Dudek[1], Piotr Arcab[1], Emilia Wdowiak[1], Paweł Matryba[2,3], Marzena Stefaniuk[2], Piotr Zdańkowski[1], Maciej Trusiak[1,*]

[1]Warsaw University of Technology, Institute of Micromechanics and Photonics, Faculty of Mechatronics, Warsaw, Poland

[2]Laboratory of Neurobiology, Nencki Institute of Experimental Biology of Polish Academy of Sciences, Warsaw, Poland

[3]Department of Immunology, Medical University of Warsaw, Warsaw, Poland

~co-first authors

*maciej.trusiak@pw.edu.pl

**Abstract**

Holotomography (HT) has revolutionized quantitative label-free 3D imaging, yet conventional lens-based implementations are fundamentally constrained in field-of-view (FOV) and imaging depth, limiting their utility for critical high-throughput applications in material and life sciences. Lensless HT (LHT) offers a promising alternative for large-volume examination, however existing approaches fail to accurately reconstruct highly scattering samples over extended depths, which remains a critical challenge in optical imaging field. Here, we introduce a gigavoxel-scale, multiple-scattering-aware LHT with a large FOV (surpassing 0.6 cm$^2$), millimeter-scale axial range and pixel level (~2.4 µm) resolution. Our approach leverages a multi-wavelength, oblique-illumination hologram reconstruction and a robust, automatic illumination angle calibration, which are necessary for precise large-volume 3D holographic reconstruction. Moreover, we propose optimization-driven multi-slice tomographic framework to accurately capture multiple-scattering effects outperforming first order Born/Rytov-based inversions. To rigorously validate our method, we reconstruct bespoke multi-layer two-photon polymerized test structure over a 1.7 mm imaging depth and 25 mm$^2$ FOV, yielding an unprecedented 3D space-bandwidth product exceeding a gigavoxel level. Furthermore, we demonstrate for the first time on-chip label-free imaging of entire 500-µm-thick tissue slice of optically-cleared mouse brain. With the proposed method, we aim to unlock powerful new capabilities for large-scale, quantitative, label-free 3D imaging across biomedicine, neuroscience, material sciences and beyond.

**Introduction**

Three-dimensional (3D) label-free optical imaging is a transformative tool in modern microscopy, enabling the visualization of biological and technical specimens with high spatial resolution while preserving their native volumetric structure[1]. Among the most powerful techniques, optical diffraction tomography (ODT)[2,3] has emerged as indispensable method for quantitative phase imaging, providing deep insights into the internal organization of transparent samples without the need for staining. ODT can be mainly realized via interferometric coding as holotomography (HT)[4] or non-interferometric coding as intensity diffraction tomography[5] or Fourier ptychographic tomography[6,7]. These 3D imaging methods have significantly advanced fields such as biomedicine, materials science, and neuroscience, allowing for the study of cellular and tissue architecture, engineered microstructures, and complex scattering media[1–7].

Despite their success, conventional implementations remain fundamentally constrained by the optical limitations of lens-based systems, where the trade-off between numerical aperture, field-of-view (FOV), and imaging depth imposes severe restrictions on practical applications. High-resolution, high-quality imaging is inherently limited to small volumes, while attempts to increase the accessible FOV often come at the cost of lower resolution[8], reduction of the illumination angles and spatial fidelity[9] or laborious and erroneous scanning and stitching[10,11].

Lensless holotomography (LHT) has recently emerged as a promising alternative, leveraging on-chip defocused intensity detection and digital holographic wavefield processing to enable tomographic imaging over large volumes without the need for bulky optical elements. By circumventing the constraints of conventional lenses, LHT holds the potential to dramatically expand the scale and versatility of tomographic imaging. LHT is a relatively new field, still in its proof-of-concept phase, with only a modest number of studies addressing its inherent challenges.

In a pioneering 2011 paper[12], Ozcan's group employed orthogonal illumination via a mechanically scanned optical fiber imaging scattering biological samples (*C. elegans*) up to ~50 µm thick. The use of an LED light source introduced coherence limitations on information content and resolution, especially for objects placed far from the sensor[13]. The 2D reconstruction lacked twin-image removal, while the 3D reconstruction did not account for multiple scattering. A portable version of this method demonstrated impressive results[14] but exhibited reduced resolution. Ozcan's group also showed a 200 µm thick mouse brain slice image[15], however the 3D effect was achieved via z-stacking and not through rigorous Fourier diffraction theorem (ODT).

Zuo et al.[16] used a color LED matrix for illumination and aimed at wide-field volumetric imaging. While enabling a large FOV (24 mm²), imaging depth remained limited to ~50 µm, likely due to the small number of angles and restricted coherence of the LED source. Berdeu et al.[17,18] developed an LHT platform based on a rotatable robotic arm and incorporated diffraction effects by applying the Fourier diffraction theorem inversion under the first-order Born approximation. However, additional image transformations (image registration, z-scaling, transverse shifts) – prone to illumination angle errors - were required, reducing imaging fidelity. The reliance on single-hologram noniterative reconstructions[17] led to strong twin-image artifacts, necessitating heavy regularization and sparsity priors that generally decrease the accuracy and resolution of 3D reconstructions of thick and dense multiply-scattering bio-structures[17,18,19]. Luo et al.[20] introduced a scan-free LHT system that uses only four laser diodes yielding a limited volume of interest (3.4 × 2.3 × 0.3 mm³) with a highly total-variation regularized (applicable to piecewise-constant samples) first-order Born reconstruction.

Recent developments in 2024 further highlight the field's rapid expansion. Zhou et al.[21] proposed an LED-based setup incorporating a deconvolution-driven tomographic reconstruction algorithm with a limited imaging volume (930 × 930 × 200 µm³). Qin et al.[22] demonstrated a single-shot portable LED-matrix-based LHT system with limited imaging depth to 50 µm. Zuo's group[22] further advanced LHT with a motion-free setup employing wavelength scanning. Despite using a broad spectral range (430–1200 nm), posing a risk of dispersion-induced errors, axial 3D Fourier space coverage remained limited.

Although these methods have demonstrated the feasibility of large scale label-free tomographic imaging on a lensless platform, they predominantly rely on direct inversion under the first-order Born or Rytov approximation[2], which is inadequate for stronger scattering in dense or thick samples[3]. These proof-of-concept approaches are typically tailored to specific cases involving sparse or weakly scattering objects, such as microbeads and thin cellular structures. In addition, LHT systems apply hologram recording in a defocused plane, which poses a major challenge when applying tilted illumination. Despite its conceptual appeal, LHT has yet to provide a comprehensive solution that delivers high-fidelity, robust, large-volume imaging with versatility required for practical deployment.

Addressing these crucial problems we introduce a gigavoxel-scale multiple-scattering-aware LHT approach for precise, large FOV and deep milimeter-scale 3D label-free phase (refraction) and amplitude (absorption) imaging. In our approach we leverage a set of novel methods: (1) robust, automated sub-degree-precision illumination angle calibration strategy, (2) 2D complex fields reconstruction from multi-wavelength inclined Gabor holograms using specialized oblique-illumination diffraction computation and (3) a reliable, optimization-driven, computationally efficient, tomographic reconstruction framework with multiple-scattering correction. We demonstrate the effectiveness of our numerically enhanced LHT framework through the successful 3D imaging of inclined phase test target, two-photon-printed bespoke volumetric calibration structure and challenging mouse brain tissue slice of unprecedented 500 µm thickness.

# Results

## Holotomographic data acquisition and calibration

Figure 1(a) illustrates the proposed experimental setup consisting of a board-level camera, the measured sample and inclined illumination (collimated fibre-coupled supercontinuum laser with tuneable wavelength). The output of the fibre is mounted on a motorized rotation stage, enabling precise angular scanning across a full 360° range. At each angular position (in total 180 projections with 2° increment), two inclined Gabor holograms are recorded – one using $\lambda_1 = 530$ nm and the other $\lambda_2 = 630$ nm illumination wavelength – for efficient phase retrieval.

Unlike lens-based HT systems limited by microscope objective's parameters, lensless configurations can reconstruct optical fields over substantially larger etendue (Supplementary Note 1). However, achieving high reconstruction accuracy over such volumes requires precise knowledge of the illumination angle ($\alpha_n$) for each $n^{\text{th}}$ collected hologram (defocused image). Any, even arcminute, angular inaccuracies lead to transverse misalignments of the propagated fields and distortions of retrieved projections, severely degrading final 3D reconstruction quality. For instance, propagating an optical field over a 1 mm distance with 40° illumination generates a 1.2 μm transverse error for just a 0.1° angle deviation.

To address this challenge, we developed SHARP – Snell-corrected Holographic Angle Recovery Protocol – a three-step lensless holotomography calibration procedure, Fig. 1. It is designed for accurate estimation of sample axial location and all illumination angles, to obtain distortion-free full-volume reconstructions, and additional correction of transverse misalignments caused by both slight angular deviations and the Snell effect due to refractive index mismatch between air, sample and camera glass cover. SHARP requires acquiring a single additional hologram under on-axis illumination (Fig. 1(b1)) and the presence of a calibration object (CO) within the imaging volume. The CO is defined as a small (10–100 μm diameter), thin (<10 μm) structure that exhibits high spatial-frequency content and is spatially isolated from other features in the sample, e.g., a dust particle or a native object feature.

In the first step of the proposed method, the on-axis hologram is numerically propagated to multiple axial planes with the use of angular spectrum method[23–25]. At each plane, we compute the DarkFocus (variance of computational darkfield gradient) sharpness metric[26] to localize the CO plane and determine its axial position $z_{co}$ relative to the camera (Fig. 1(b2)). In the second step, we analyze each $n^{\text{th}}$ hologram acquired under oblique illumination (Fig. 1(c1)) to estimate the corresponding $x$ and $y$ illumination angle components $\alpha_{x_n}$ and $\alpha_{y_n}$. For each dataset, the CO sharpness is evaluated using the DarkFocus metric at the holograms propagated at previously identified $z_{co}$ plane, across a range of candidate illumination angles centered around the nominal values obtained from the acquisition system (blue points in Fig. 1(c2)). In this case, holograms are propagated by precise off-axis diffraction numerical modelling method, described in detail in Supplementary Note 3. The final illumination angle is determined by locating the maximum of a second-order polynomial fit to the computed sharpness metric values.

Ideally, numerically propagating each hologram to the CO plane using the corresponding calibrated angle should yield reconstructions where the CO appears in-focus at the same transverse location. In practice, small residual angular errors and the refractive shift caused by the sample glass slide, coverslip and protective cover glass on top the camera sensor (due to the Snell effect) introduce transverse displacements, visualized in Fig. 1(d). Therefore, in the third step, we compute the residual transverse shift ($\xi_{x_n}$, $\xi_{y_n}$) between the CO reconstructions obtained under $1^{st}$ and $n^{th}$ illumination angles using a cross-correlation approach (Fig. 1(d1))[27]. This compensates for cumulative misalignments and ensures that the CO is correctly registered across all illumination angles. For the planes located far from the CO plane, residual angular calibration errors can still introduce transverse misalignments, potentially degrading lateral resolution. This effect can be mitigated by repeating the calibration procedure for additional axial planes. However, it is important to note that for all datasets presented in this work, no significant resolution loss was observed across volumes up to 10×5×2 mm³.

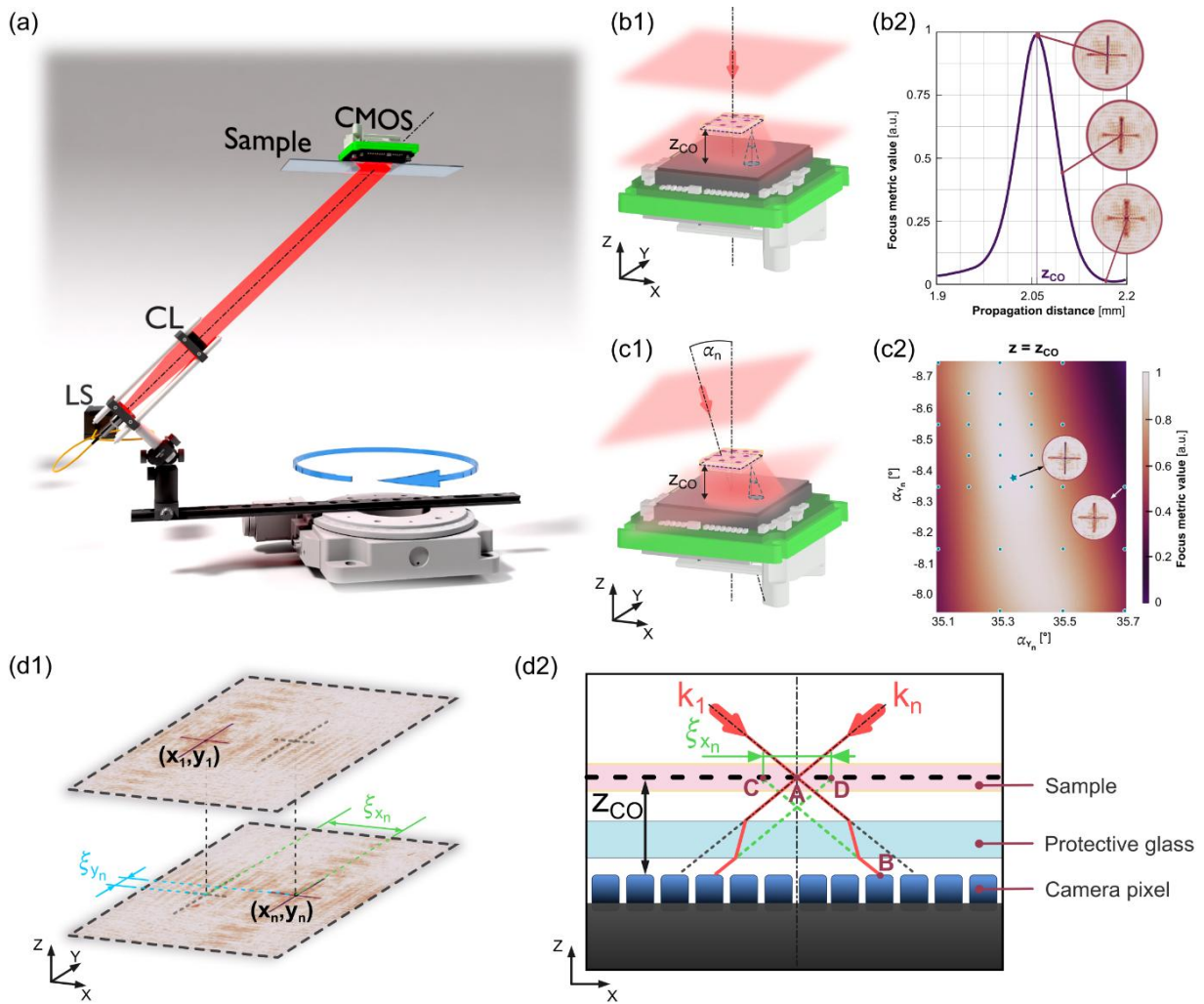

***Fig. 1. Lensless holotomograph design and calibration principle.*** *(a) Schematic illustration of the lensless gigapixel-scale quasi-isotropic holotomography setup. A laser source (LS) emits (sequentially with $\lambda_1 = 530$ and $\lambda_2 = 630$ nm) a coherent beam that is collimated by a collimating lens (CL) to illuminate the sample at an oblique angle. LS is mounted on a precision rotation stage, enabling full angular diversity for cone shape tomographic illumination pattern. A high-resolution CMOS sensor captures the diffracted light without the use of imaging lenses, enabling lensless gigavoxel-scale reconstructions. (b1) System configuration for acquiring the on-axis*

*calibration-only hologram. (b2) DarkFocus metric values computed for the CO across various propagation distances under normal illumination ($\alpha = 0°$). (c1) System configuration for oblique illumination. (c2) DarkFocus metric values evaluated at different candidate illumination angles, with the axial distance fixed at $z_{co}$; blue dots represent the calculated metric values, color-coded image represents the 2D polynomial fitted to calculated values and blue star represents the found illumination angle (location of the maximum of a polynomial). (d1) Lateral misalignment between reconstructions of the CO for $1^{st}$ and $n^{th}$ illumination angles, quantified as $\xi_{x_n}, \xi_{y_n}$. (d2) Illustration of the Snell effect influence onto the system an object at position A illuminated with wavevector $k_1$ appears at position B on the camera sensor due to refraction by the cover glass. Reconstruction of $1^{st}$ hologram under $k_1$ yields a shifted image CO at point C, while reconstruction of $n^{th}$ hologram under $k_n$ leads to CO image at D, offset by $\xi_{x_n}$ from C.*

**Lensless multiple-scattering-aware tomographic large-volume reconstruction**

Following the SHARP calibration procedure, the collected holograms $I_{\lambda_1}(x, y)$ and $I_{\lambda_2}(x, y)$ undergo preprocessing – they are shifted by $\xi_x$ and $\xi_y$ to correct for transverse misalignments and normalized by dividing them by their mean value. Next, the complex amplitudes in the camera plane, $C_n(x, y)$, are retrieved for each $n^{th}$ illumination angle. For this, we are using our modified off-axis multi-wavelength Gerchberg–Saxton[16,28,29] (GS) phase retrieval algorithm, tailored for volumetric sample imaging with oblique illumination (Algorithm S1 in Supplementary Note 4).

For efficient large-volume gigavoxel-scale robust 3D reconstruction, we propose a novel algorithm called Scattering-Optimized Lensless Volume Estimator – SOLVE. Our method effectively captures the multiple-scattering effects that have long hindered LHT's application to dense, axially extended complex samples. Firstly, an initial estimate of the 3D complex amplitude distribution is obtained using a modified filtered backpropagation (FB) algorithm[30] (Algorithm S2 in Supplementary Note 5) for the user defined range of propagation distances and the measured volume axial sampling. The final 3D distribution ($U$) is then refined using the proposed iterative reconstruction algorithm.

Each iteration of the SOLVE algorithm (Algorithm S3 in Supplementary Note 6) begins with filtering the 3D complex volume $U$ by applying a physical constraint that the normalized object amplitude must be less than or equal to 1 – meaning the object can absorb but not generate light. This step improves the convergence and stability of the reconstruction process. Next, for each illumination angle $n$, the multi-plane volume $U$ is propagated to the camera plane using a multi-slice beam propagation method (BPM) model[31]. The propagation between individual slices of $U$ is realized with the use of specialized off-axis shift-preserving AS method, while propagation between last slice of $U$ and camera is realized with shift-suppressed AS. Both AS variants are tailored to specific LHT conditions and are thoroughly described and motivated in Supplementary Note 3, offering additional insights in comparison to previous works[22,32]. The difference between the estimated field at the camera and the measured complex amplitude (retrieved from oblique GS) is then calculated. After computing a full set of differences, a 3D correction volume $\Delta U$ is reconstructed using the FB method. This $\Delta U$ is subtracted from the current $U$ with an empirical learning rate, resulting in the updated estimate.

**Experimental verification**

To experimentally evaluate the system's transverse and axial resolution in large-volume, we reconstructed a custom phase resolution target with etched features down to 2 μm width, Fig. 2. The target was inclined (~25°) to imitate 3D object and enable axial sectioning verification. Phase reconstructions revealed clear resolution of 3 μm-wide elements, while 2 μm elements remained unresolved, consistent with the 2.4 μm sensor pixel pitch. Slight artifacts observed in axial cross-sections are attributed to the limited illumination angle range. Figure 2(f) displays the minimum phase projection of the 3D volume along the axial direction full-field overview of high-fidelity imaging.

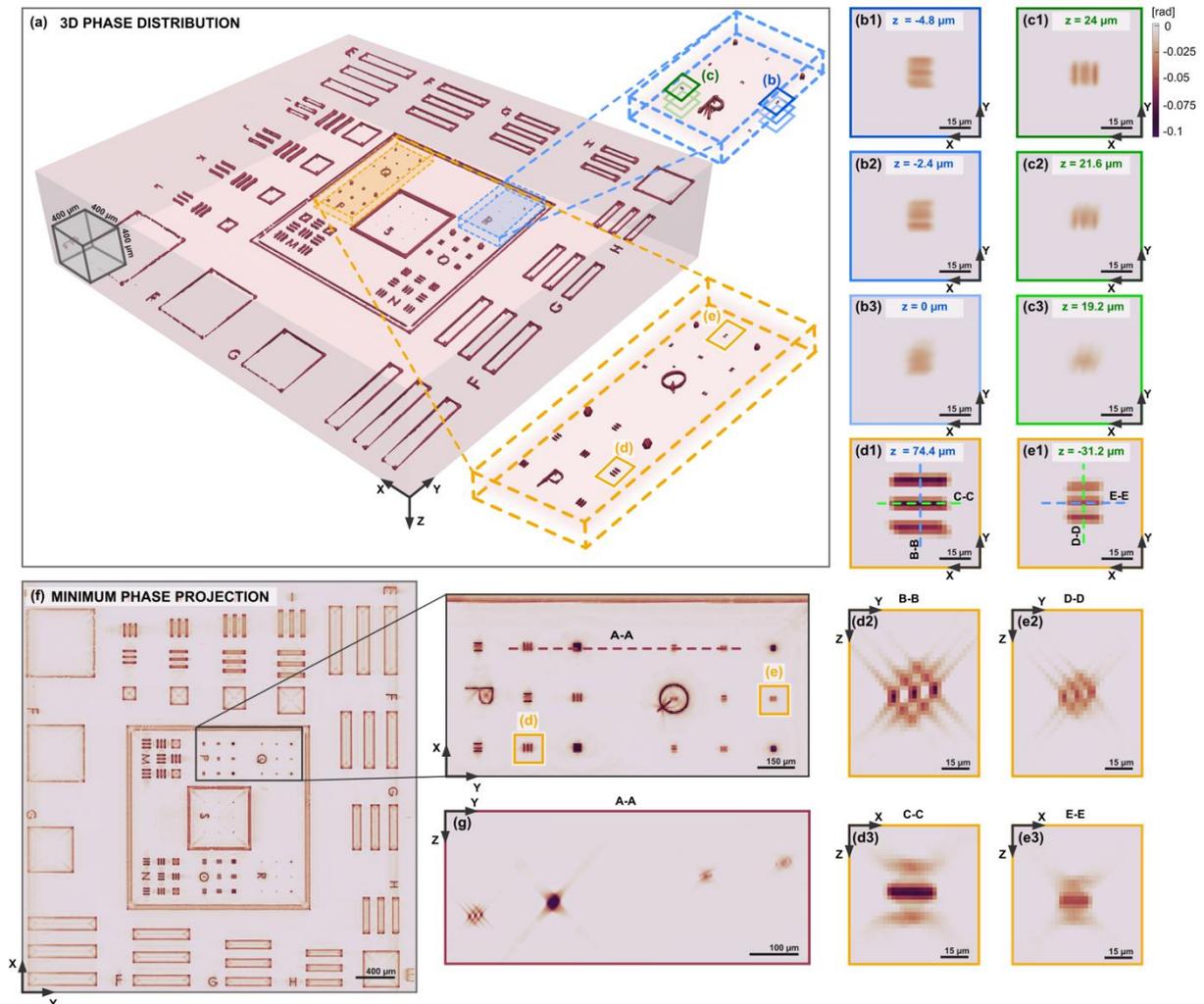

***Fig. 2. 3D phase reconstruction of a resolution target.*** *A custom phase resolution target (etched 125 nm features on Borofloat 33 glass) was imaged at a tilt of 25°. (a) 3D rendering of the reconstructed phase distribution. (b1) – (b3) Transverse phase slices of 3 μm-wide features (group R) along x axis, in focus (b1) and at different axial planes (b2), (b3). (c1) – (c3) Transverse phase slices of 3 μm-wide features (group R) along y axis, in focus (c1) and at different axial planes (c2), (c3). (d1) – (d3) Transverse, coronal, and sagittal representation of 6 μm elements from*

*group P. (e1) – (e3) Transverse, coronal, and sagittal representation of 4 μm elements from group Q. (f) Minimum phase projection of the entire volume reveals global structure but shows hollow centres in larger features (e.g., groups E–H), consistent with known limitations of in-line holography in recovering low-frequency phase components. (g) Sagittal cross-section through group P and Q elements. Mild reconstruction artifacts above/below structures are attributed to the limited angular coverage (missing cone). See Visualization 1 for a full 3D rendering of the phase test reconstruction.*

Figure 3 presents transverse and axial cross-sections of the P-group elements from the tilted resolution phase target shown in Fig. 2, reconstructed using the FB method (Figs. 3(a), 3(b)) and the proposed SOLVE algorithm (Figs. 3(c), 3(d)), for both single-wavelength (Figs. 3(a), 3(c)) and multi-wavelength (Figs. 3(b), 3(d)) data. In the single-wavelength case, reconstructions were obtained using the same FB and SOLVE procedures, but with intensity-only input fields $C(x,y)$ estimated as the square root of the recorded intensity holograms $I_{\lambda_1}$, without any GS-based phase retrieval. All methods show out-of-plane artifacts (e.g., highlighted with an arrow in Fig. 3(b3)), which are primarily due to the limited range of illumination angles[33–35]. However, in the reconstructions obtained with the proposed SOLVE algorithm, these artifacts are significantly suppressed confirming the improved axial accuracy capability of the SOLVE method. Moreover, the use of an additional wavelength, combined with the oblique-illumination GS-based phase retrieval, effectively reduces twin-image and background-related artifacts. This leads to improved phase background uniformity and overall tomographic reconstruction quality.

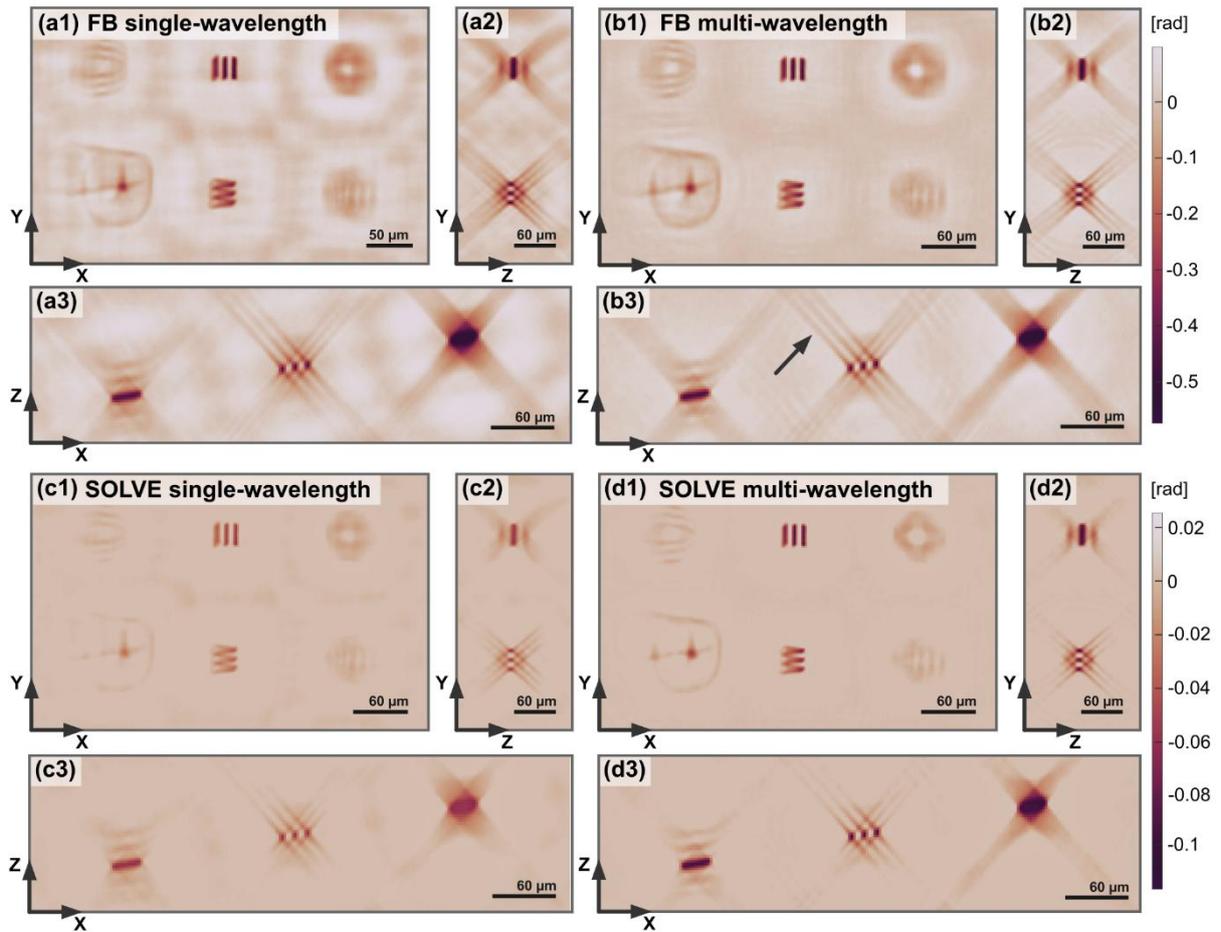

*Fig. 3. Impact of the proposed SOLVE algorithm and multi-wavelength reconstruction on 3D imaging accuracy. Selected XY (1), YZ (2) and XZ (3) slices through 3D phase reconstructions based on: (a) intensity-only holograms and (b) GS retrieved complex fields under single-scattering model (FB method). (c) intensity-only holograms and (d) GS retrieved complex fields under multiple-scattering model (SOLVE method). The use of complex-field modelling and multiple-scattering propagation notably improves image fidelity, sharpness, and suppression of artifacts, particularly along the axial dimension. Further comparisons of the FB and SOLVE algorithms applied to a dense, multiple-scattering object are provided in Supplementary Note 2.*

Figure 4 presents the reconstruction of our bespoke 3D test sample composed of multiple phase objects arranged at different axial positions. Results demonstrate the capability of the proposed method to perform optical sectioning and distinguish multiple two-dimensional phase structures positioned at different depths, while effectively minimizing defocus contributions from out-of-focus planes. Successful 1.7 mm deep imaging of 25 mm$^2$ FOV experimentally verifies gigavoxel-scale space-bandwidth product of our method.

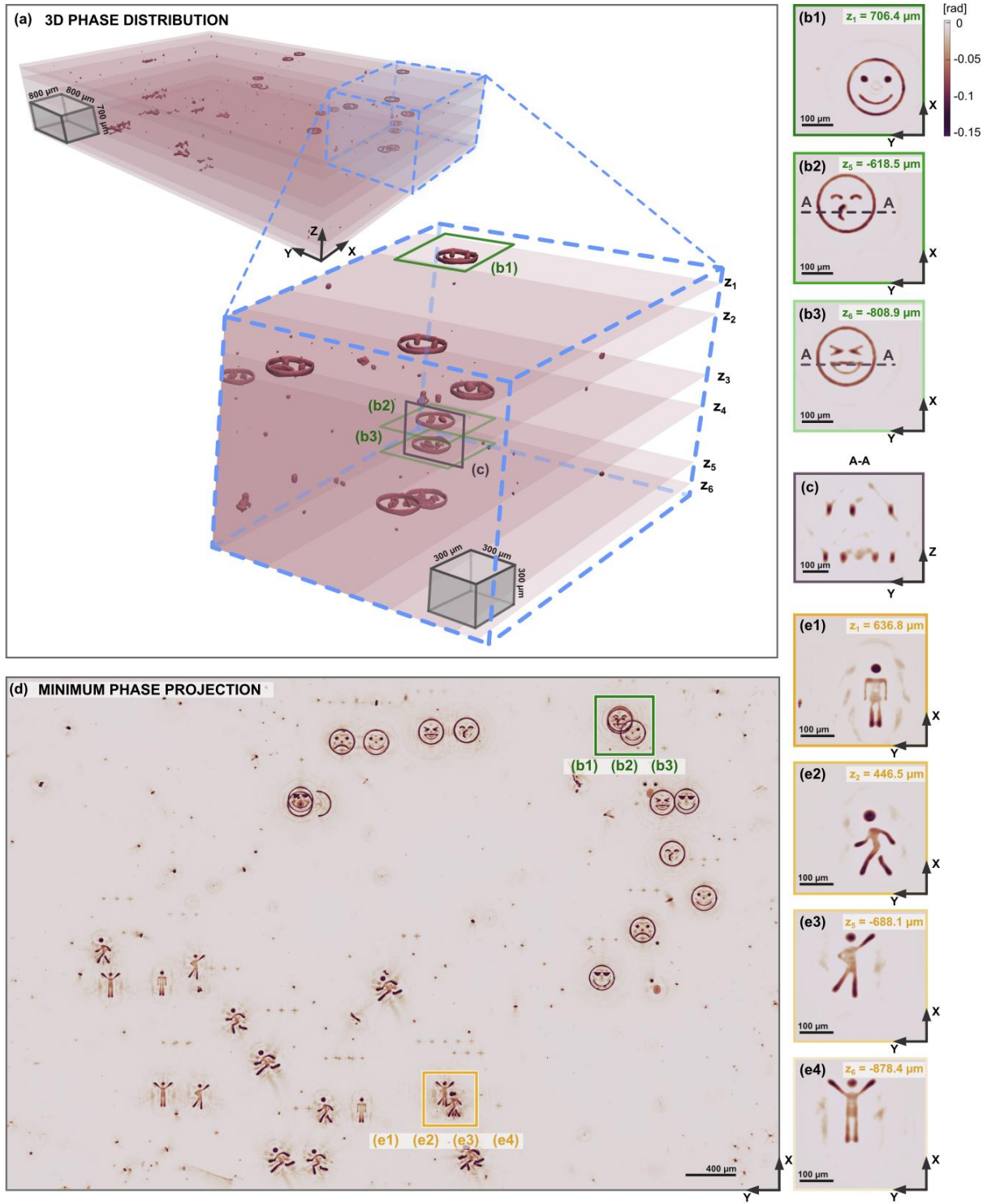

**Fig. 4. Phase reconstruction of a bespoke multilayer 3D phase test sample.** *(a) 3D rendering of the reconstructed phase distribution. (b1) – (b3) Transverse phase slices showing three distinct emoji patterns located at the same lateral position, separated along the z-axis. (c) Longitudinal cross-section through two emoji patterns from (b2) and (b3). (d) Minimum phase projection in the transverse plane. (e1) – (e4) Transverse regions of four different stick figures, each located at a different axial plane. A discrepancy in the axial position of the $z_1$ plane can be observed between (b1), where z1 = 706.4 µm, and (e1), z1 = 636.8 µm. This difference arises from a slight*

*unintentional tilt introduced to the 3D test sample during the measurement process. Visualization 2 shows a full 3D rendering of our reconstruction.*

To demonstrate the capability of our system in imaging thick, morphologically-rich biological tissues, we performed a 3D reconstruction of a 500 μm-thick mouse brain section optically cleared using the CUBIC protocol[36,37]. By omitting a step when samples are cut into thin sections it is possible to crucially preserve long projections between distant brain regions. The reconstructed volume – Fig. 5 – reveals distinct morphological features depending on the imaging contrast. Blood vessels, anatomical boundaries, and small structures are highlighted differently in amplitude and phase reconstructions, emphasizing the complementary nature of both modalities. The panels in Fig. 5(b1–b6) and (d1–d6) present detailed minimum phase and intensity projections across localized regions and axial depths. Each panel integrates several adjacent z-slices centered around key structural planes to preserve continuity and detail. Vascular features are clearly visible in phase, e.g., Fig. 5(b1–b2), and largely absent in amplitude, Fig. 5(d1–d2). Figure 5(d1) shows an imaging artifact unique to amplitude modality, not observed in the corresponding phase projection. Notably, the anterior commissure olfactory (aco) is well defined in the amplitude projection at a specific axial plane (d5), while the caudoputamen (CPu) appears distinctly only in phase (b5), at Z = 76.7 μm, and is not visible at neighboring depths, Fig. 5(b4, b6), underscoring the high depth resolution of our approach. For anatomical reference, panel Fig. 5(c) includes a Nissl-stained histological section and corresponding annotated map from the Allen Mouse Brain Atlas[38].

The mouse brain section had in-plane dimensions of 9.5 mm × 7 mm, slightly exceeding the FOV of our system in the y-direction. As a result, portions of the diffracted light fell outside the camera sensor area (see Visualization 3 for full holographic sequence); however, each region of the sample was covered in at least three-quarters of collected projections. This redundancy enabled 3D reconstruction of the entire section without the need for mechanical stitching or sample/camera motion. The sample was immersed in CUBIC-R1 solution during measurement to reduce refractive index mismatches.

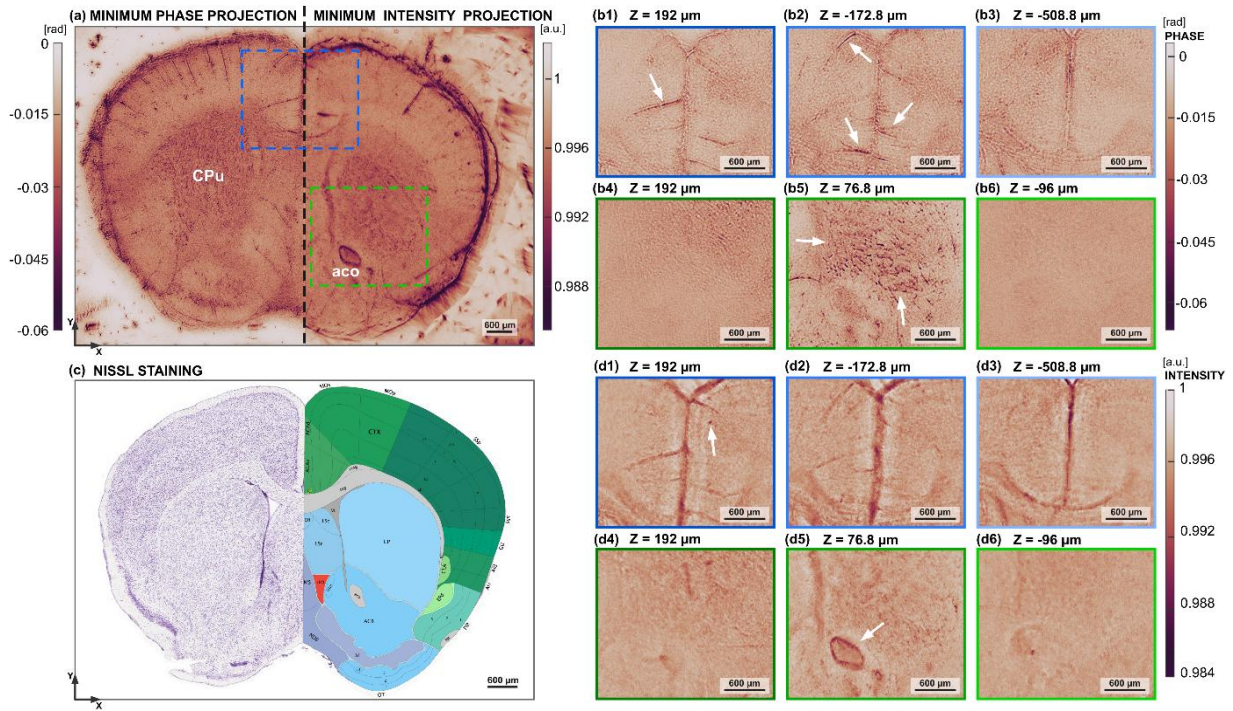

*Fig. 5. Reconstruction of a 500 µm thick optically cleared (CUBIC protocol) mouse brain section.* (a) Minimum phase and intensity projections of the left and right hemisphere, respectively, in the transverse plane indicating CPu – caudoputamen and aco – anterior commissure olfactory. (b1) - (b6) Transverse phase slices showing fine phase anatomical details (white arrows), separated along z-axis, in areas marked by dashed blue and green rectangles in (a). (c) Corresponding Nissl-stained section (on the left) next to the brain map from the Allen Brain Atlas (on the right)[38]. (d1) - (d6) Transverse intensity slices showing fine amplitude anatomical details (white arrows), separated along z-axis, in areas marked by dashed blue and green rectangles in (a). Visualization 3 shows the complete tomographic sequence of input holograms.

## Discussion

The LHT systems are generally limited in resolution by the pixel size, however demonstrate overall superior tomographic imaging capability, as their transfer functions - analyzed in detail in Supplementary Note 1 - are much less anisotropic than for the lens-based HT. The LHT, with approx. 40 degrees cone illumination, thus provides better, i.e., more isotropic, coverage of the 3D object spectrum suggesting quasi-isotropic hardware-limited (pixel size – 2.4 µm in our case) resolution in XYZ (transversal and axial). The FOV in LHT is, in principle, limited only by the physical dimensions of the camera sensor (13.2 × 8.8 mm in our case). Diffraction patterns originating near the edges of the sample may, however, fall outside the camera sensor area for certain illumination directions due to defocused imaging geometry. Despite this fundamental information reduction, the achievable effective FOV remains significantly larger than that of conventional lens-based HT systems (up to around 7000 times larger comparing to commercial Nanolive system).

Standard multiple-scattering tomographic reconstruction algorithms[31,39–42] iteratively improve the 3D reconstruction by minimizing the discrepancy between the experimentally acquired data and the model predictions using gradient descent minimization and are generally computationally expensive. Proposed SOLVE approach improves the 3D reconstruction by iteratively backpropagating the optical field corrections, given by a difference between the multi-slice model prediction of the optical field and the actual acquired data. A conceptually strategy has previously been employed only in lens-based systems[43–46]. This alternative error-reduction mechanism for multiple-scattering tomography has been demonstrated to be effective and computationally efficient[47] laying foundation for our SOLVE method. SOLVE is implemented in space domain thus it eliminates the need for direct filling of the 3D Fourier space and – crucially for large-volume imaging - avoids the related interpolation errors. Additionally, space-domain implementation enables tackling the problem of deterioration of the reconstruction accuracy in the peripherical regions of the tomographic volume that is typical for Rytov-based direct inversion algorithms[48,49]. Moreover, it offers the potential to accommodate arbitrarily shaped illumination beams – particularly spherical beams, useful in compact LHT.

By addressing fundamental optical and computational challenges, our LHT approach enables high-fidelity, large-volume 3D imaging with multiple-scattering correction, previously unattainable without complex optics and laborious scanning approaches. The integration of scalable hardware with the new reconstruction and calibration algorithms establishes a versatile platform suitable for diverse applications across biology, medicine, and materials science. We anticipate that this framework may inspire further innovations in lens-free volumetric microscopy and expand accessibility to high-throughput label-free advanced 3D imaging.

## Methods

**Experimental setup.** Proposed experimental setup consists of a board-level CMOS camera (ALVIUM 1800 U-2050 m mono Bareboard, 2.4 µm pixel size, 5496 × 3672 resolution), the measured sample (positioned around 2 mm above the camera sensor), a collimating lens CL (f = 200 mm) and a coherent illumination light source LS (Supercontinuum laser, NKT Photonics SuperK EVO), delivered to the system with a single-mode optical fiber (Thorlabs P1-S405-FC-2, core diameter: ~3 µm), see Fig. 1(a). The output of the fiber is mounted on a motorized rotation stage (Standa 005460), enabling precise angular scanning of the illumination beam across a full 360° range. Illumination azimuth is easily adjustable and was generally set between 35 and 45 degrees to ensure quasi-isotropic XYZ reconstruction, while maintaining relatively large FOV. During the dataset acquisition, the rotation stage steps in 2° increments resulting in N=180 angular positions. At each position, two holograms are recorded – one using $\lambda_1 = 530$ nm and the other $\lambda_2 = 630$ nm illumination wavelength – resulting in a total of 360 images per dataset. Additionally, one on-axis hologram (with 530 nm wavelength) is registered only for angle calibration purposes.

**2PP bespoke volumetric test target fabrication.** Two-photon polymerization target consists of three high-precision coverslips (each 170 μm thick, with ne = 1.5255 ± 0.0015), resulting in six discrete test planes ($z_1 - z_6$ in Fig. 2). Each plane contains distinct patterns of 30 μm thick emojis and stick figures, fabricated using a two-photon polymerization technique (Photonic Professional GT2, Nanoscribe GmbH) on both sides of coverslip (0.17 mm thick). The structures were designed in the DeScribe software and composed of IP-S photoresist (refractive index around 1.515 when fully polymerized), selected for its nearly pure phase characteristics due to its polymeric nature. The coverslips are spaced using 430 μm thick spacers and immersed in ne = 1.518 Nikon Type N immersion oil. Additional clean coverslips are placed on the top and bottom of the sample bringing the total thickness of the 3D sample to approximately 1.7 mm (largest distance between two printed structures is 1.37 mm).

**Mouse brain sample preparation.** Brain was collected from the adult C57Bl6 mouse. Mouse was anesthetized using overdose of sodium pentobarbital. Next it was perfused as described previously[50]. Brain tissue was collected, postfixed overnight in 4oC. Next, it was embedded in 4% agarose to facilitate cutting. Brain was cut using vibratome (Leica) into 500 μm sections and stored for the analysis in PBS solution in 4oC. Tissue optical clearing The CUBIC-R1 solution was prepared according to a published protocol[36]. A clearing solution was prepared as a mixture of 25% (mass fraction) urea (Sigma-Aldrich), 25% (mass fraction) N; N; N0; N0-tetrakis(2-hydroxypropyl)ethylenediamine (Sigma-Aldrich), and 15% (mass fraction) Triton X-100 (Sigma-Aldrich) dissolved in a distilled water. For tissue optical clearing, the agarose was removed from the brain slices, which were then submerged in an excess (around 5 ml) of CUBIC-R1 solution in a custom-made Petri dish to minimize the potential effects of solution evaporation.

**Acknowledgements**

Funded by the European Union (ERC, NaNoLens, Project 101117392). Views and opinions expressed are however those of the author(s) only and do not necessarily reflect those of the European Union or the European Research Council Executive Agency (ERCEA). Neither the European Union nor the granting authority can be held responsible for them.

**Authors contribution**

M.T. conceived and supervised this project and provided funding for the project. M.T., P.Z., M.R., J.W., J.D. and P.A. designed the research. M.R. and J.D. developed calibration method. M.R. and J.W. developed the reconstruction algorithms. J.W. designed the propagation algorithms and performed the 3D transfer function analysis. J.D. and M.R. performed the data preprocessing and reconstruction. P.A. built the tomographic system. P.A. and J.D. acquired the data. E.W. designed and prepared the 3D printed samples. E.W., J.W., and P.Z. generated the manuscript figures. P.M. provided and prepared mouse brain sample. M.S. provided the biological analysis of brain reconstruction. M.T., M.R. and J.W. wrote the manuscript and all authors commented on the manuscript.

**Competing interests**

The authors declare no competing interests.

**Data availability**

The raw data supporting the findings of this study are available upon resonable request.

**Code availability**

The algorithms used in this study are provided as part of the Supplementary Information.

**Supplementary Information**

Supplementary Information is available for this paper in Supplementary Document 1. It includes Supplementary Notes 1–6, Supplementary Figures S1–S6, and Supplementary Algorithms S1–S3.

# Supplementary Document 1

## Supplementary note 1: Theoretical transfer functions, resolution and field of view analysis

This part analyses the tomographic imaging capability of conventional lens-based holotomography (HT) and lensless holotomography (LHT) by comparing their 3D transfer functions (TF). The TFs presented in Fig. S1 were obtained using identical system parameters: a conical illumination scenario with an incidence angle of $\alpha = 40°$, a light wavelength of $\lambda = 0.54$ µm, and a refractive index of the surrounding medium of $n_i = 1$.

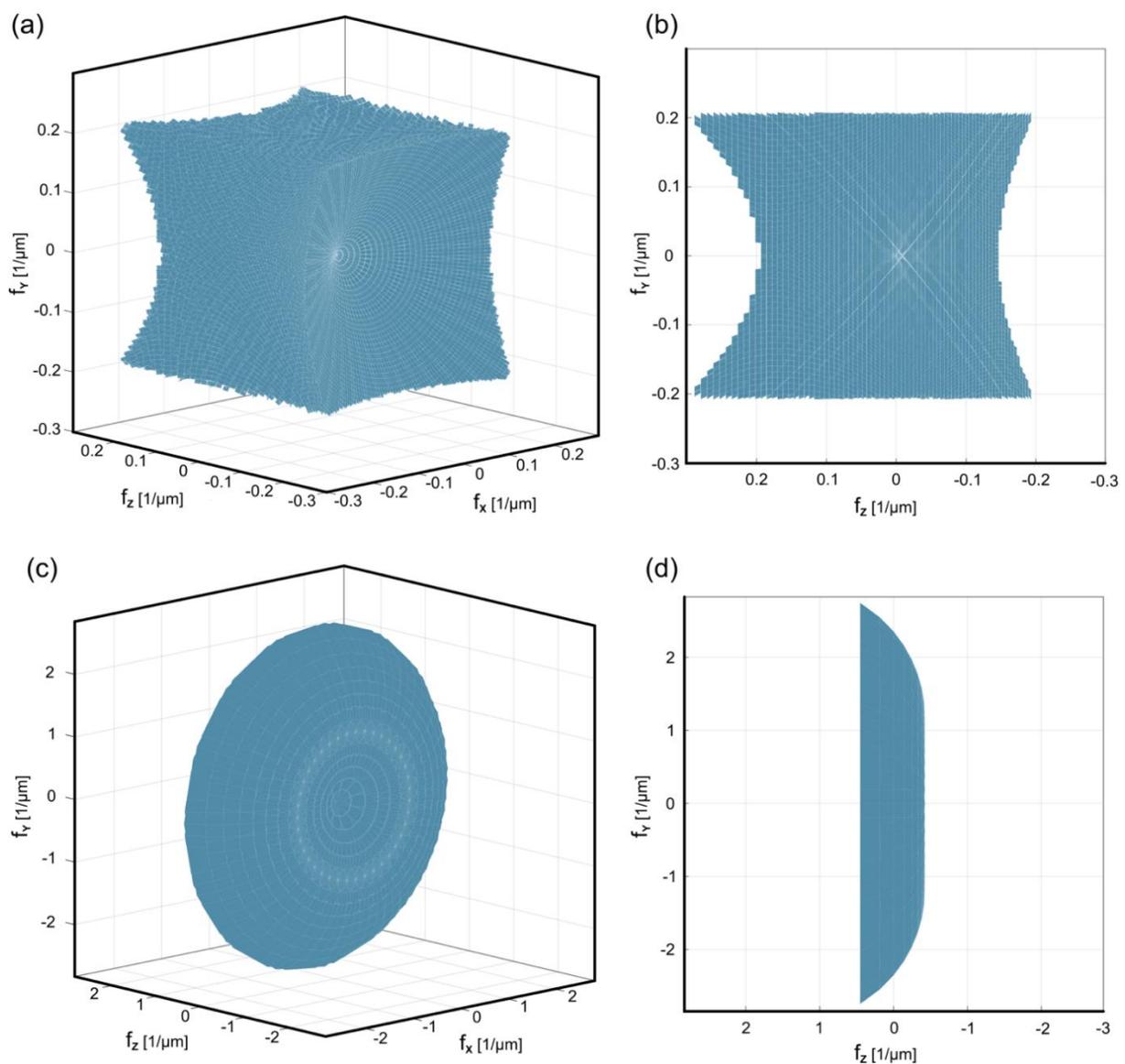

Fig. S1. Two 3D renderings of transfer functions for tomographic imaging with lensless (a, b) and conventional lens-based systems (c, d). The applied parameters are as follows: a conical illumination scenario with an incidence angle of 40°, light wavelength of $\lambda = 0.54$ µm, and a refractive index of the

*surrounding medium of $n_i$ = 1. The transverse resolution is limited by the pixel size ($\Delta_x$ = 2.4 µm) in the lensless system (a) and by the numerical aperture (NA = 0.85) in the lens-based setup (c). Lensless tomographic imaging provides a favorable quasi-isotropic imaging condition, as evidenced by the axial depth comparison between (b) and (d).*

In the typical lensless tomography system, the transverse resolution is limited by the pixel size $\Delta_x$; therefore, the spatial frequency limit of 3D imaging in the transverse direction is given by $f_x^{LHT} = 1/(2\Delta_x)$ (here $\Delta_x$ = 2.4 µm thus $f_x^{LHT}$ = 0.208 $\mu m^{-1}$). Notably, in lensless systems, this frequency limit is independent of the illumination angle. This also means that steep illumination angles, which are required for tomographic reconstruction, can be applied even in relatively low-resolution systems. In contrast, for lens-based systems, the maximum transferred spatial frequency $f_x^{HT}$ depends on both the numerical aperture (NA) of the imaging system and the applied illumination angle: $f_x^{HT} = (NA + n_i sin\alpha)/\lambda$. Moreover, in lens-based system the illumination angle must fall within the NA-related acceptance angle: $\alpha < \alpha_{NA} = \mathrm{asin}(NA/n_i)$, which imposes the requirement for high-NA optics.

For our comparative analysis (Fig. S1), we assumed an illumination angle of 40°, which matches the value used in our experimental setup. To apply the same angle in a lens-based tomographic system, the numerical aperture must satisfy NA>sin(40°)≈0.643. In our simulation, we assumed NA = 0.85, which corresponds to an optics used in a widely used, commercial lens-based HT system by Nanolive. Under these conditions $f_x^{HT} = 2.76 \mu m^{-1}$. It is important to note that high NA must be accompanied by large transverse magnification (60x for Nanolive), which results in a small FOV (85 x 85 µm for Nanolive). In contrast, the lensless system supports steep illumination while maintaining small (unity) magnification enabling tomographic imaging over significantly larger FOV (sensor size dependent, however easily up to 5 x 5 mm), allowing for whole-slide tissue slice and cell culture imaging without the time-consuming and quality-decreasing need for scanning and stitching.

Lastly, the LHT system demonstrates overall superior tomographic imaging capability, as its TF is much less anisotropic than for the lens-based HT, see Fig. S1. Notably, both considered systems (lensless and lens-based) utilize the same limited-angle tomographic architecture, which results in analogical missing cone problem, i.e., the conical region centered around the $f_z$ axis where object information is missing. However, as seen from comparison of the axial width of LHT and HT in Figs. S1(b) and S1(d), the lensless imaging system provides better, i.e., more isotropic, coverage of the 3D object spectrum. In more detail, the extension of lensless tomography TF for positive and negative axial frequencies is given by the following limits:

$$f_z^{LHT} = \begin{cases} \frac{n_i}{\lambda}\cos\alpha - \sqrt{\left(\frac{n_i}{\lambda}\right)^2 - \left(\frac{\sqrt{2}}{2\Delta_x} + \frac{n_i}{\lambda}\sin\alpha\right)^2} & \text{for positive } f_z \\ \frac{n_i}{\lambda}\cos\alpha - \sqrt{\left(\frac{n_i}{\lambda}\right)^2 - \left(\frac{\sqrt{2}}{2\Delta_x} - \frac{n_i}{\lambda}\sin\alpha\right)^2} & \text{for negative } f_z \end{cases}. \quad (1)$$

The corresponding half-width is given by:

$$HW_z^{LHT} = 0.5\left[\sqrt{\left(\frac{n_i}{\lambda}\right)^2 - \left(\frac{\sqrt{2}}{2\Delta_x} - \frac{n_i}{\lambda}\sin\alpha\right)^2} - \sqrt{\left(\frac{n_i}{\lambda}\right)^2 - \left(\frac{\sqrt{2}}{2\Delta_x} + \frac{n_i}{\lambda}\sin\alpha\right)^2}\right], \quad (2)$$

which for the given parameters results in $HW_z^{LHT}$ = 0.257 µm⁻¹. Notably, in this case, the axial extent of 3D transfer function is even slightly larger than the traverse extent ($f_x^{LHT}$ = 0.208 µm⁻¹).

For comparison, the axial half-width of lens-based HT transfer function (assuming conical illumination scenario) is given by:

$$HW_z^{HT} = \frac{n_i}{\lambda}\left(\sin\frac{\alpha_{NA}}{2}\right)^2, \quad (3)$$

which for the considered parameters gives $HW_z^{HT} = 0.438$ µm⁻¹. This means that the transverse extent of TF is approximately 6 times larger than its axial dimension: $f_x^{HT} = (NA + n_i \sin\alpha)/\lambda = 2.76$ µm⁻¹.

We demonstrated that the extended axial imaging depth is accompanied by relatively strong depth discrimination (tomogram axial resolution), enabled by the unique off-axis defocused imaging mechanism of lensless holographic microscopy.

## Supplementary note 2: Comparison of LHT reconstruction algorithms for multiple-scattering objects

Figure S2 presents a comparison of the 3D reconstruction of a Colosseum model structure using a two-photon polymerization system. The STL model[1] was translated into printing coordinates and fabricated using the IP-S photoresist (Nanoscribe). The structure measures 500 × 400 × 134 µm³ (length × width × height) and is characterized by a refractive index of 1.515. It was fabricated on a high-precision 170 µm-thick coverslip and immersed in Nikon Type N immersion oil (refractive index $n_e$ = 1.518). Figures S2(a)–S2(d) show 3D visualizations of reconstructions obtained using the filtered backpropagation (FB) method (Figs. S2(a), S2(b)) and the proposed SOLVE method (Figs. S2(c), S2(d)), for both single-wavelength (Figs. S2(a), S2(c)) and multi-wavelength (Figs. S2(b), S2(d)) datasets. The visualizations were generated using Tomviz software, with threshold cutoff values set to 0.1165 rad (Fig. S2(a)), 0.1663 rad (Fig. S2(b)), 0.0146 rad (Fig. S2(c)), 0.0228 rad (Fig. S2(d)) – all thresholds were set manually to highlight the structure of the object while minimizing visible missing-cone related artifacts. Figure S2(e) shows the ground truth STL model used for two-photon polymerization printing of the test structure.

As shown, reconstructions based on single-wavelength data suffer from prominent twin-image artifacts, particularly visible in the empty central region of the Colosseum structure (highlighted with blue ellipse in Figs. S2(a) and S2(c)). In contrast, the use of multi-wavelength data significantly suppresses these artifacts, enabling more accurate separation of object information from the background during thresholding.

Further comparison between the FB and SOLVE reconstructions reveals marked differences in reconstruction fidelity. The FB-based reconstruction displays structural heterogeneity: certain regions appear substantially thicker than their SOLVE counterparts (green regions in Figs. S2(b) and S2(d)), while other parts present in the SOLVE result are completely missing from the FB reconstruction (yellow regions in Figs. S2(b) and S2(d)). These discrepancies highlight the higher fidelity of the SOLVE reconstruction, as its greater uniformity is more consistent with the fact that the printed object had a homogeneous refractive index.

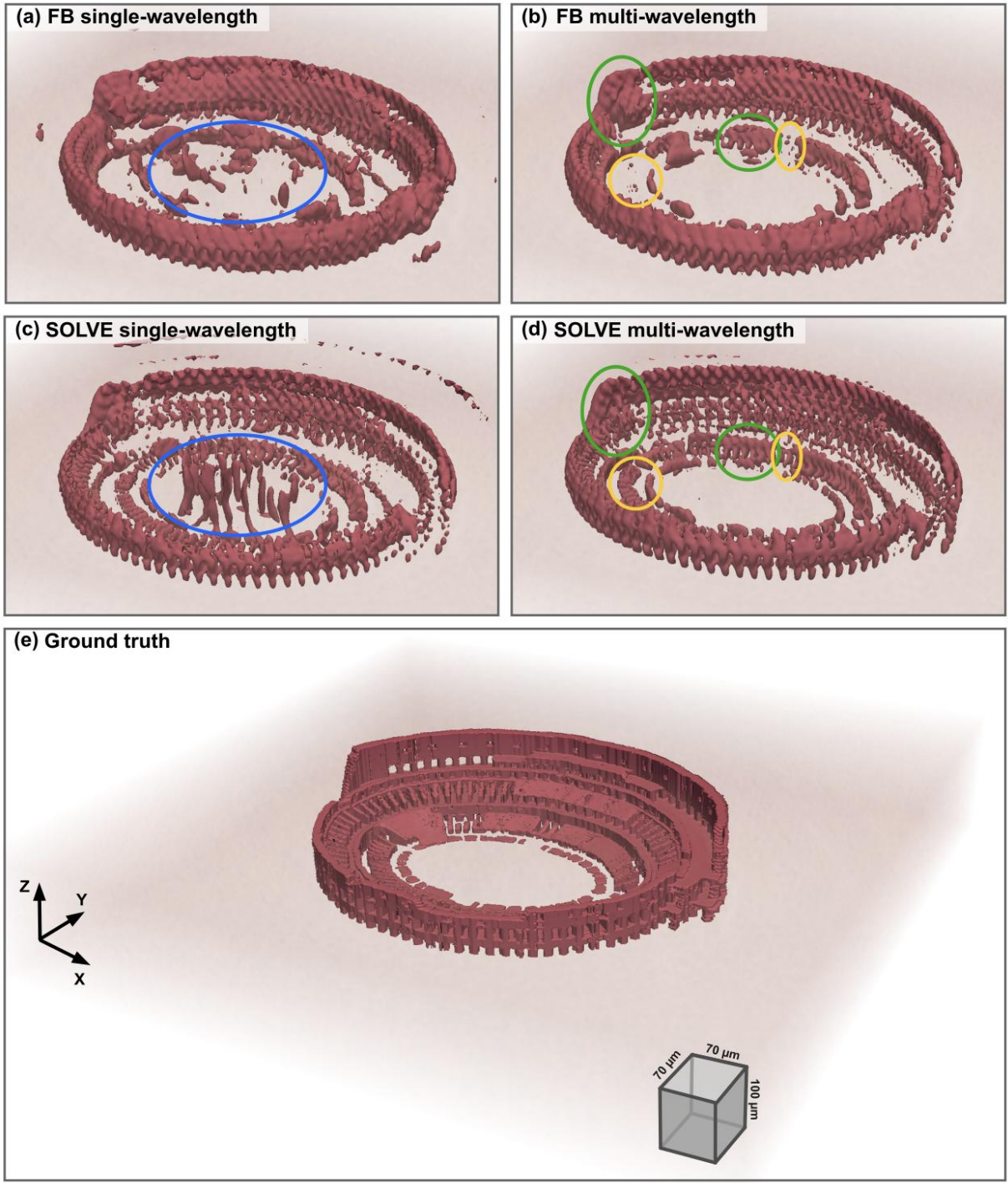

*Fig. S2. 3D phase reconstructions of Colosseum model fabricated with two-photon polymerization visualizing the impact of the proposed SOLVE algorithm and multi-wavelength reconstruction on 3D imaging accuracy. Reconstructions based on: (a) intensity-only holograms and (b) GS retrieved complex fields under single-scattering model (FB method). (c) Intensity-only holograms and (d) GS retrieved complex fields under multiple-scattering model (SOLVE method). (e) Ground truth STL model[1].*

# Supplementary note 3: Tailored angular spectrum propagation algorithm for lensless holotomography

A critical aspect of holographic reconstruction in lensless microscopy involves backpropagating the optical field captured at the camera plane to the object plane. In lensless microscopy systems with non-unity magnification – where the curvature of the illumination wavefront cannot be neglected – specialized numerical refocusing algorithms with adaptive sampling intervals are employed[2-4]. In our system, however, unity magnification is used employing collimating lens.

Nonetheless, a significant challenge arises due to the steep illumination angles required by the tomographic reconstruction algorithm. Our reconstruction algorithm is based on a widely adopted method utilizing optical field decomposition into plane waves, known as the Angular Spectrum (AS)[5]. The first step of the AS method is to compute the two-dimensional Fourier transform of the optical field $U_{z=0}$ in the source plane $z=0$:

$$\widetilde{U}_{z=0}(f_x, f_y) = \iint U_{z=0}(x,y) exp[-i2\pi(f_x x + f_y y)] dx dy, \tag{4}$$

which corresponds to the decomposition of the signal into plane waves. In the above expression, $f_x$ and $f_y$ represent the spatial frequencies of the object beam. Next, the product of the optical field spectrum and the free-space propagation transfer function is calculated:

$$\widetilde{U}_z(f_x, f_y) = \widetilde{U}_{z=0}(f_x, f_y) exp\left[i2\pi z \sqrt{\left(\frac{n_i}{\lambda}\right)^2 - f_x^2 - f_y^2}\right], \tag{5}$$

where λ is a light wavelength, $n_i$ denotes refractive index of surrounding medium, and $z$ is the propagation distance. The last step is inverse Fourier transform:

$$U_z(x,y) = \iint \widetilde{U}_z(f_x, f_y) exp[i2\pi(f_x x + f_y y)] df_x df_y. \tag{6}$$

Figure S3 illustrates the AS algorithm with simulation-based example. Since our focus is on propagation method for limited angle tomography, we consider a tilted propagation direction $(\alpha_x, \alpha_y)$ that corresponds to Fourier frequencies ($f_{ix}=n_i \sin\alpha_x/\lambda$, $f_{iy}=n_i \sin\alpha_y/\lambda$) (in our example, for simplicity, we assumed $\alpha_y = 0$). The input optical field is modelled as the product of the illumination field, here, tilted plane wave $\exp[i2\pi(f_x x + f_y y)]$ (Fig. S3(e)), and the sample contribution, which in this case is centrally located circular amplitude modulation (Fig. S3(a)). We assume following parameters: light wavelength λ=0.54 μm, propagating in a medium with refractive index $n_i$=1 at angle $\alpha x$=40⁰ on a distance of $z$=30 μm. The optical field is discretised with 1024x1024 grid and sampling interval is equal to 0.3λ but later in this section we will also consider sampling with larger interval, i.e., Δ=2.4 μm. The choice of the parameters value is motivated by the condition in the experimental system of our lensless tomogram.

The amplitude and phase of the source optical field $U_{z=0}(x,y)$ is displayed in Fig. S3(a) and S3(e), respectively. The off-axis character of $U_{z=0}$ can be seen from its phase tilt in Fig. S3(e). Figures S3(b) displays the amplitude of the spectrum $\widetilde{U}_{z=0}$ of the source field. During the AS algorithm, $\widetilde{U}_{z=0}$ is multiplied with the propagation transfer function (Figures S3(c) and S3(f)), and then inverse Fourier transformed to produce the output field $U_z(x,y)$ (Figures S3(d) and S3(g)). It can be seen that the tilted propagation direction introduces numerical errors in the form of diffraction artifacts at the boundary of the computational matrix (artificial vertical lines in Fig. S3(d)). These errors result from the use of the fast Fourier transform, which assumes periodic boundary conditions and leads to discontinuities at the image edges due to the phase ramp associated with the off-axis propagation direction.

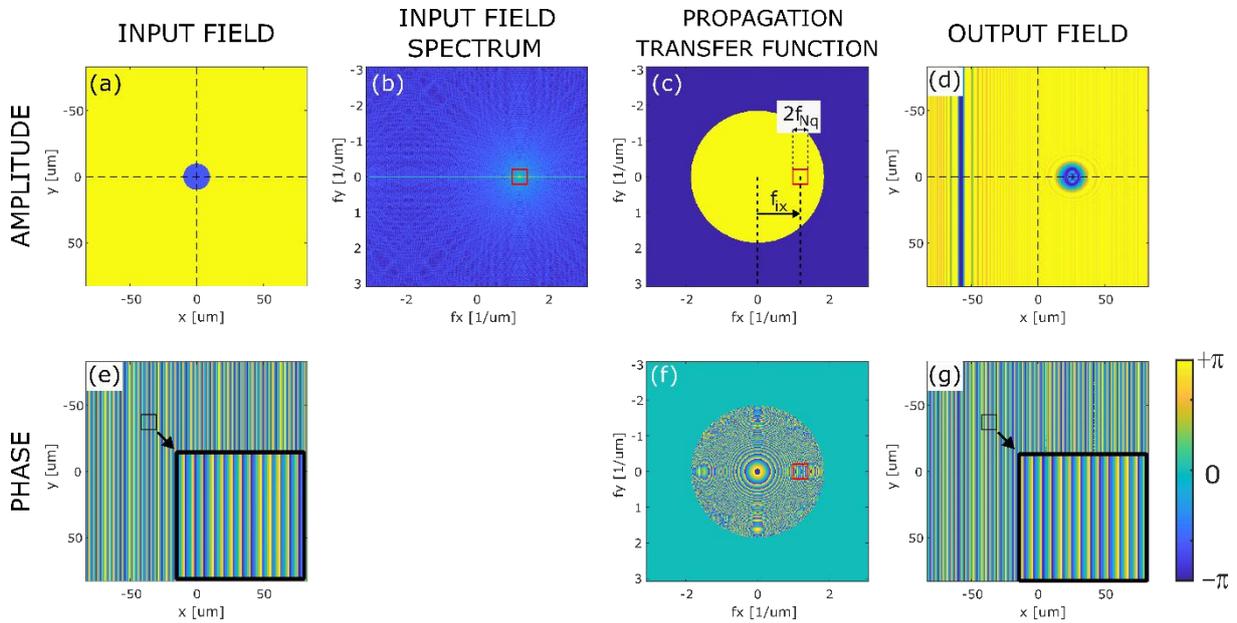

*Fig. S3. Visualization of the AS algorithm: amplitude (top row) and phase (bottom row) of (a,e) the input optical field, (b) the spectrum of the input optical field, (c,f) the propagation transfer function, and (d,g) the output optical field. Results obtained for: wavelength λ = 0.54 µm, illumination angle $α_x$ = 40°, propagation distance z = 30 µm. The red rectangle indicates the Fourier frequency range that is transferred by lensless microscope with a pixel size of 2.4 µm.*

The red rectangle in Fourier spectrum (Fig. S3 (b),(c),(f)) denotes the Fourier frequency range of a single object wave that is accessible with lensless microscopy, where the limiting factor for resolution is typically the camera's pixel size Δ (here we assumed Δ=2.4 µm). The resolution limitation mechanism is as follows – the pixel size must be small enough to allow the recording of interference fringes formed between the wave scattered by the object and the undisturbed illuminating wave. This implies that, unlike in classical microscopy, the frequency limitation applies to spatial frequencies shifted by the illumination frequency vector ($f_{ix}$, $f_{iy}$). In other words, the system transmits spatial frequencies within a rectangular aperture (red rectangle in Fig. S3(c)) localized around the spatial frequency ($f_{ix}$, $f_{iy}$):

$$\begin{cases} |f'_x| = |f_x - f_{ix}| < f_{Nq} \\ |f'_y| = |f_y - f_{iy}| < f_{Nq} \end{cases}, \qquad (7)$$

where Nyquist frequency $f_{Nq}$ =1/(2Δ). Notably, this is a unique feature of the lensless microscopy. In the lens-based system, resolution is typically limited by the numerical aperture of the optical setup and is always cantered around the zero Fourier frequency of the object wave.

In summary, AS algorithm, when applied to lensless holotomography, faces two main challenges. First, inclined illumination introduces boundary artifacts. Second, the relevant Fourier components of the object waves are shifted away from the center of the computational matrix. As a result, the algorithm requires significantly greater computational resources – namely, smaller pixel sizes and a larger number of pixels – to accurately capture the shifted spectrum.

As it will be shown, both of these issues can be effectively resolved by performing computations in shifted spatial frequencies coordinates, defined as $(f'_x, f'_y) = (f_x - f_{ix}, f_y - f_{iy})$. The modified algorithm is illustrated with Figure S4. First, the input object wave $U_{z=0}(x,y)$ is multiplied by a conjugate illumination plane wave, resulting in a modified source field $U^{lf}_{z=0}(x,y)$ with the phase tilt removed (Fig. S4(e)). This operation effectively shifts the object wave's spectrum to the centre of the computational matrix (superscript *lf* stands for low-frequency shifted wave):

$$U^{lf}_{z=0}(x,y) = U_{z=0}(x,y)exp[-i2\pi(f_{ix}x + f_{iy}y)]. \qquad (8)$$

Next, the standard AS steps - Fourier transformation, multiplication with transfer function, and inverse Fourier transform – are carried out in the shifted spatial frequency coordinates (Fig. S4 (b),(c),(f)):

$$\widetilde{U}_{z=0}(f'_x, f'_y) = \iint U^{lf}_{z=0}(x,y)exp[-i2\pi(f'_x x + f'_y y)]dxdy, \qquad (9)$$

$$\widetilde{U}_z(f'_x, f'_y) = \widetilde{U}_{z=0}(f'_x, f'_y)exp\left[i2\pi z\sqrt{\left(\frac{n_i}{\lambda}\right)^2 - f'^2_x - f'^2_y}\right], \qquad (10)$$

$$U^{lf}_z(x,y) = \iint U_z(f'_x, f'_y)exp[i2\pi(f'_x x + f'_y y)]df'_x df'_y. \qquad (11)$$

The final result is a propagated beam with the carrier frequency removed and boundary artifacts significantly reduced (Figs. S4(d) and S4(g)). Notably, this feature of the modified AS have been already discussed in the context of application to lens-based holotmography[6]. If needed, the original carrier frequency can be reintroduced by multiplying $U^{lf}_z$ by the illumination plane wave:

$$U_z(x,y) = U_z^{lf}(x,y)exp[i2\pi(f_{ix}x + f_{iy}y)]. \tag{12}$$

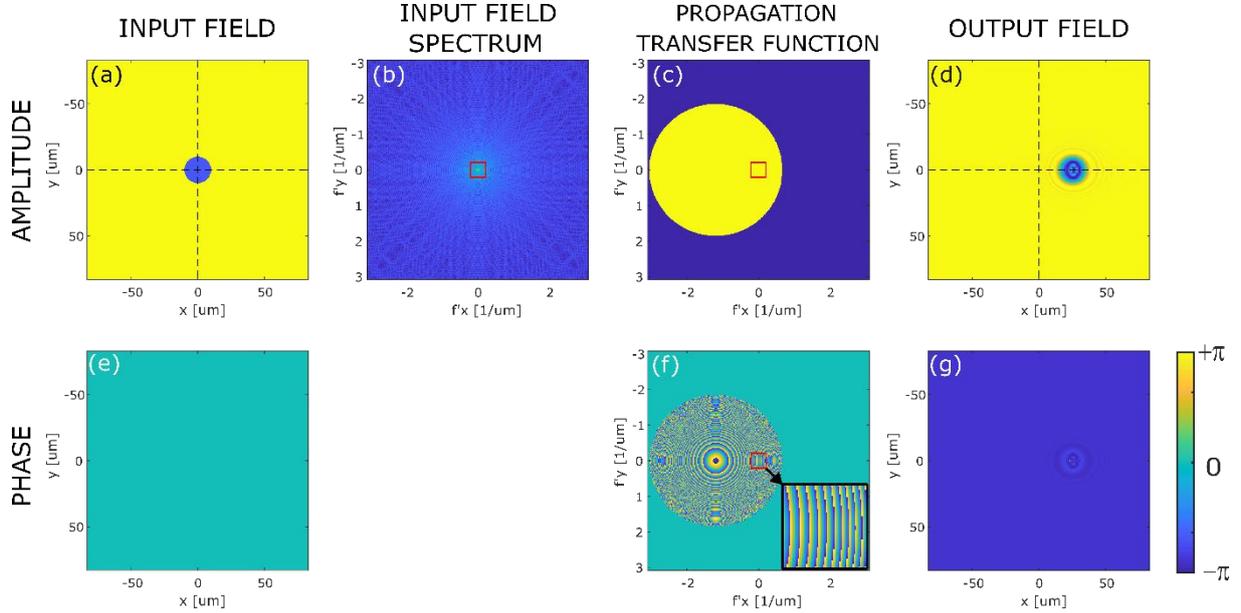

*Fig. S4. Visualization of the modified Shift-Preserving, Tilt-Removed Angular Spectrum (SP-TR-AS) algorithm: amplitude (top row) and phase (bottom row) of (a,e) the input optical field, (b) the spectrum of the input optical field, (c,f) the propagation transfer function, and (d,g) the output optical field. Results obtained for: wavelength λ = 0.54 µm, illumination angle $α_x$ = 40°, propagation distance z = 30 µm. The red rectangle indicates the Fourier frequency range that is transferred by lensless microscope with a pixel size of 2.4 µm. In SP-TR-AS, the tilt of the plane wave carrier is numerically removed, which supresses the boundary errors and enables using course sampling of the input signal.*

The modified algorithm addresses the problem of border errors and ensures that the spectral region of interest is aligned with the center of the computational matrix, allowing propagation to be performed with fewer computational resources (that is smaller spectral bandwidth and thus larger sampling interval Δ of the signal). However, one challenge remains. As shown in Figure S4(f), the phase of the propagation transfer function varies rapidly within the central region (region of interest marked with red frame). This means that to properly sample the transfer function, a high sampling frequency of the spectrum is required. A sampling interval of the spectrum is given by $Δ_{fx}$=1/N/Δ, where N is the number of pixels. Therefore, to properly sample the transfer function, a large field of view is necessary. It is important to stress that this effect becomes very significant over large propagation distances (e.g., millimetre-scale backpropagation during hologram reconstruction in lensless microscopy), because the curvature of the propagation transfer function's phase distribution, $2\pi z\sqrt{\left(\frac{n_i}{\lambda}\right)^2 - f_x'^2 - f_y'^2}$, is inversely proportional to the propagation distance. This makes aliasing-free sampling of the propagation transfer function particularly challenging.

The described problem can be overcome by removing the phase tilt from the transfer function. The required tilt value is strictly related to the illumination vector and happens to be equal to transverse object wave shift vector $(x_i, y_i)$ that results from the tilted propagation direction. This shift can be calculated from the geometry. It is helpful to change Cartesian representation of the propagation angle $\alpha_x, \alpha_y$ to spherical coordinates:

$$\alpha = \cos^{-1}\left(\frac{f_{iz}}{\frac{n_i}{\lambda}}\right) = \cos^{-1}\left(\frac{\sqrt{\left(\frac{n_i}{\lambda}\right)^2 - f_x^2 - f_y^2}}{\frac{n_i}{\lambda}}\right), \quad (13)$$

$$\theta = \tan^{-1}\left(\frac{f_y}{f_x}\right), \quad (14)$$

where $\alpha$ is a tilt with respect to the optical axis and $\theta$ is azimuth of illumination. The total shift in a transverse direction after propagation on a distance $z$ is given by:

$$r = z \cdot \tan\alpha. \quad (15)$$

The transverse shift vector can be calculated from:

$$\begin{aligned} x_i &= r\cos\theta \\ y_i &= r\sin\theta \end{aligned}. \quad (16)$$

The final AS-based propagation algorithm that is dedicated for the resources-efficient object wave backpropagation in lensless holotomography is described with Equations 8-12, where Eq. (10) is replaced with the following formula:

$$\tilde{U}_z(f_x', f_y') = \tilde{U}_{z=0}(f_x', f_y')exp\left[i2\pi z\sqrt{\left(\frac{n_i}{\lambda}\right)^2 - f_x'^2 - f_y'^2}\right]exp[-i2\pi(f_x x_i + f_y y_i)]. \quad (17)$$

Notably, this solution has already been proposed[7] and applied for LHT[8]. However, here we fully justify when and why this choice is optimal for LHT and support this claim with simulation-based analysis.

Removing the linear phase component of the transfer function [Eq. (17)] reduces the sampling requirements of the signal (see slowly changing transfer function phase in Fig. S5(f)). Therefore, the challenging off-axis long-distance propagation can be accurately performed using minimal sampling, i.e., evaluation in shifted spectral coordinates enables relatively large pixel size that is typical for lensless microscopy, whereas the transfer function phase tilt removal enables using relatively small size of the computational matrix (number of pixels $N$). Additionality, in accordance with the Fourier shift theorem, the phase tilt removal in the Fourier domain results in the propagated optical field being represented in a shifted spatial coordinate system ($x-x_i$, $y-y_i$). As a result, despite the tilted beam propagation, the diffraction shadow of the object occupies the same area of the computational matrix (compare position of the circle in the source (Fig. S5(a)) and destination plane (S5(d))). Therefore, removing the phase tilt not only enables optimal sampling conditions for propagation but also eliminates the sample's transverse shift, simplifying the data preprocessing pipeline for lensless holotomography.

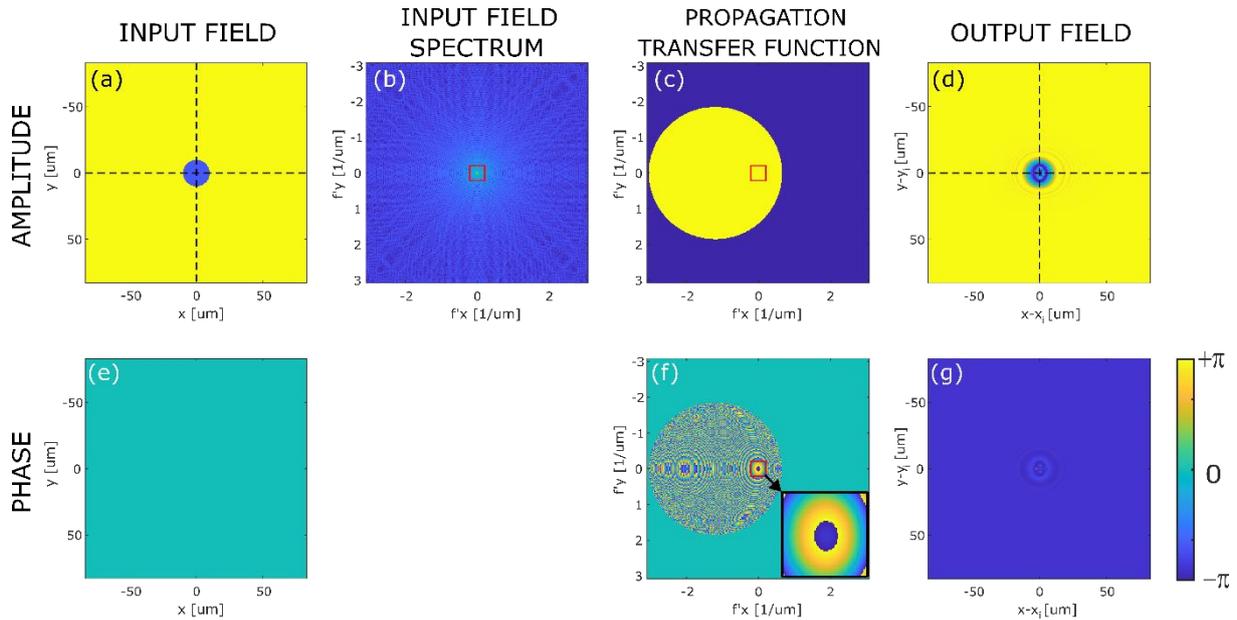

*Fig. S5. Visualization of the modified Shift-Suppressed, Tilt-Removed Angular Spectrum (SS-TR-AS): amplitude (top row) and phase (bottom row) of (a,e) the input optical field, (b) the spectrum of the input optical field, (c,f) the propagation transfer function, and (d,g) the output optical field. Results obtained for: wavelength λ = 0.54 μm, illumination angle $α_x$ = 40°, propagation distance z = 30 μm. The red rectangle indicates the Fourier frequency range that is transferred by lensless microscope with a pixel size of 2.4 μm. SS-TR-AS removes the carrier tilt and also numerically suppresses the associated transverse shift. This allows for a more compact computational domain (courser sampling and smaller field of view required).*

In summary, this section describes two modified versions of the Angular Spectrum (AS) algorithm, tailored for the numerical modelling of the propagation of tilted optical fields. The first method, defined by equations 8–12, is an angular spectrum propagation technique in which the tilt of the plane wave carrier is numerically removed, while the resulting transverse shift of the optical field is preserved. We refer to this method as Shift-Preserving, Tilt-Removed Angular Spectrum (SP-TR-AS). This approach mitigates boundary errors and enables propagation with coarser sampling of the computational grid.

The second method, defined by equations in order 8, 9, 13-17, 11, also removes the carrier tilt but additionally suppresses the resulting transverse shift. This allows for a more compact computational domain and improves sampling efficiency. In particular, it enables accurate and aliasing-free long-distance propagation using coarse sampling and a small field of view. We refer to this method as Shift-Suppressed, Tilt-Removed Angular Spectrum (SS-TR-AS).

In the proposed LHT reconstruction scheme we employed both SP-TR-AS and SS-TR-AS propagations. SP-TR-AS method was used for propagation on short (z=Δ) distances within the reconstructed volume, as the preservation of the transversal shift during propagation is necessary for volumetric reconstruction, and the short propagation distance does not introduce aliasing issues. SS-TR-AS method was employed for challenging off-axis propagation on larger distances (between camera and sample volume), which facilitate proper sampling of the propagation transfer function. Notably, the applied data processing pipeline—including data calibration, phase retrieval, and the tomographic reconstruction algorithm—assumes working with a low-frequency-shifted field representation (with the carrier plane wave related to the tilted illumination removed). Therefore, we omitted steps 8 and 12 in our calculations.

## Supplementary note 4: Multi-wavelength oblique illumination Gerchberg-Saxton phase retrieval for LHT data

Algorithm S1 implements a novel, computationally efficient multi-wavelength variant of the Gerchberg-Saxton algorithm. It is specifically designed for three-dimensional reconstruction under oblique illumination conditions, incorporating the captured intensities constraints and Ewald's sphere theory. The algorithm takes as input two intensity images, acquired with $\lambda_1$ and $\lambda_2$ wavelength for $n^{th}$ illumination angle ($\alpha_{x_n}, \alpha_{y_n}$). It begins with initial guess, that the complex field at camera plane for first wavelength ($C_{\lambda_1 n}$) is equal to the square root of collected intensity image $I_{\lambda_1 n}$. Next, the iterative procedure begins, in which the complex field $C_{\lambda_1 n}$ is transformed to the $C_{\lambda_2 n}$ space, next the obtained $C_{\lambda_2 n}$ field is updated with collected $I_{\lambda_2 n}$ image and transformed back to $C_{\lambda_1 n}$ space. Iteration finishes with the $C_{\lambda_1 n}$ update with $I_{\lambda_1 n}$ image.

The mentioned above transform between general $U_{\lambda_1, z=0}$ and $U_{\lambda_2, z=0}$ fields is described with following steps:

(1) Fourier transform $U_{\lambda_1, z=0}$ according to Eq. (9), which results in $\widetilde{U}_{\lambda_1, z=0}$,
(2) Multiply $\widetilde{U}_{\lambda_1, z=0}$ with the SS-TF-AS transfer function (Eq. (15)) to propagate at $+z_{co}$ distance to the sample plan: $\widetilde{U}_{\lambda_1, z_{co}}$,
(3) Remaining in the Fourier domain, transfer to $\lambda_2$ domain using the Ewald sphere theory, which results in $\widetilde{U}_{\lambda_2, z_{co}}$,
(4) Multiply $\widetilde{U}_{\lambda_2, z_{co}}$ with the SS-TF-AS transfer function (Eq. (15)) to propagate at $-z_{co}$ distance: $\widetilde{U}_{\lambda_2, z=0}$,
(5) Inverse Fourier transform of $\widetilde{U}_{\lambda_2, z=0}$ (Eq. (A8)): $U_{\lambda_2, z=0}$.

It is important to note that the employed cross-wavelength transformation method offers significant computational saving as it requires only two Fourier transforms (step 1 and 5) for each iteration. Notably, conventional multi-wavelength GS approaches[9-11] applies four Fourier transformations for each iteration – two Fourier transforms for propagation to the sample plane, in which the wavelength rescaling is achieved in space domain by direct multiplication of phase by the wavelength ratio $\varphi_{\lambda 2} = \varphi_{\lambda 1} \frac{\lambda_1}{\lambda_2}$, and then two Fourier transforms to propagate back to the camera plane. In our method, the wavelength rescaling is performed entirely in the Fourier domain, which significantly improves computational efficiency. As a result, the total computational cost is only two Fourier transforms plus the computation of two propagation kernels and one wavelength rescaling kernel (step 3, Eq. (21) - explained later), which can be precomputed at the start of the GS algorithm.

Step (3) of our method is based on Wolf's theory[12], which forms the theoretical foundation for tomographic reconstruction in holotomography and thus accounts for thick character of the sample. According to Wolf's theory, the optical field scattered by a 3D sample encodes information about the Fourier components of the sample's 3D scattering potential spectrum, $\tilde{O}$, which lie on a spherical cap. The radius of this cap is determined by the ratio of the surrounding refractive index to the wavelength of light, $\frac{n_i}{\lambda_{1,2}}$, and its position depends on the illumination direction. Specifically, the cap intersects the origin of the coordinate system and is oriented normal to the illumination direction.

In our GS method, we record two in-line holograms at different wavelengths for each illumination direction. Consequently, each angular projection corresponds to a pair of Ewald caps with different radii that intersect at the origin and are tangent to each other at that point, both oriented orthogonally to the illumination direction. This is illustrated in Fig. S6, which shows 9 pairs of Ewald caps corresponding to 9 illumination directions evenly distributed over a +/-40° angular range and two wavelengths (530 nm and 630 nm, denoted by green and red lines, respectively). As seen in Fig. S6, due to the relatively narrow range of transverse Fourier frequencies—a characteristic of lensless holographic tomography (HT) compared to lens-based HT—the two Ewald caps are closely spaced. In our method, we neglect this slight spatial offset and assume that the Ewald caps occupy the same volume. Under this approximation, we derive a quantitative relationship between the two in-line holograms, since both capture information from the same region of the scattering potential. This enables us to define a transformation between the two wavelength domains.

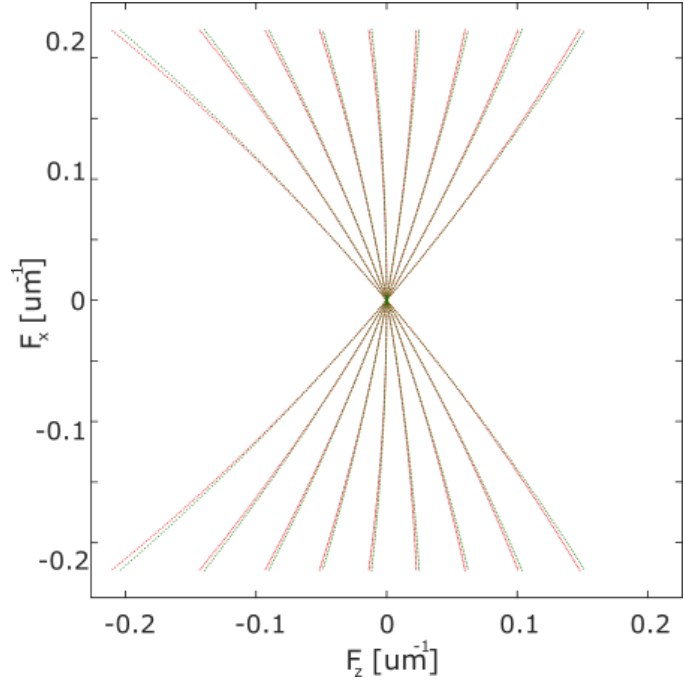

*Fig. S6. Ewald spheres corresponding to two wavelengths (530 nm and 630 nm for red and green lines, respectively) and 9 illumination angles uniformly distributed over a ±40° angular range.*

In order to establish this transform, let us first recall the first-order Born approximation from Wolf's theory, which defines the quantitative relationship between the scattered optical field and the Fourier spectrum of the scattering potential:

$$\widetilde{U}_{\lambda,z_{co}}(F_x, F_y) = \frac{\left(\frac{2\pi}{\lambda}\right)^2}{2i2\pi f_z} \widetilde{O}(F_x, F_y, F_z). \tag{18}$$

In the above formula $(F_x, F_y, F_z)$ are Fourier coordinates of the 3D spectrum. The transverse coordinates are related to the spatial frequencies of the optical fields by: $F_x = f_x - f_{ix} = f'_x$, $F_y = f_y - f_{iy} = f'_y$, where the illumination vector $f_{ix} = \frac{n_i}{\lambda}\sin\alpha_x$, $f_{iy} = \frac{n_i}{\lambda}\sin\alpha_y$ (see Sec. Tailored angular spectrum propagation algorithm for lensless holotomography). The axial Fourier coordinates are defined by $F_z = f_z - f_{iz}$, where $f_z = \sqrt{\left(\frac{n_i}{\lambda}\right)^2 - f_x^2 - f_y^2}$ and $f_{iz} = \sqrt{\left(\frac{n_i}{\lambda}\right)^2 - f_{ix}^2 - f_{iy}^2}$. Now, if we disregards the difference in location of Ewald caps due to change in light wavelength, we can claim that:

$$\frac{\left(\frac{2\pi}{\lambda_1}\right)^2}{2i2\pi f_z(\lambda_1)} \cdot \widetilde{U}_{\lambda 2,z_{co}} = \frac{\left(\frac{2\pi}{\lambda_2}\right)^2}{2i2\pi f_z(\lambda_2)} \cdot \widetilde{U}_{\lambda 1,z_{co}}, \tag{19}$$

which, after simplification, provides the cross-wavelength transform equation:

$$\widetilde{U}_{\lambda 2,z_{co}} = \left(\frac{\lambda_1}{\lambda_2}\right)^2 \frac{f_z(\lambda_1)}{f_z(\lambda_2)} \cdot \widetilde{U}_{\lambda 1,z_{co}}, \tag{20}$$

which, in practise, is computed as:

$$\widetilde{U}_{\lambda_2,z_{co}} = \left(\frac{\lambda_1}{\lambda_2}\right)^2 \frac{\sqrt{\left(\frac{n_i}{\lambda_1}\right)^2 - \left(f_x' + \frac{n_i}{\lambda_1}\sin\alpha_x\right)^2 - \left(f_y' + \frac{n_i}{\lambda_1}\sin\alpha_x\right)^2}}{\sqrt{\left(\frac{n_i}{\lambda_2}\right)^2 - \left(f_x' + \frac{n_i}{\lambda_2}\sin\alpha_x\right)^2 - \left(f_y' + \frac{n_i}{\lambda_2}\sin\alpha_x\right)^2}} \cdot \widetilde{U}_{\lambda_1,z_{co}}, \qquad (21)$$

where $(f_x', f_y')$ denote the spatial frequency coordinates resulting from the Fast Fourier Transform of optical fields $U_{\lambda_{1,2},z_{co}}$ (it is assumed that $U_{\lambda_{1,2},z_{co}}$ has removed phase tilt related to inclined illumination, see explanation in Sec. Tailored angular spectrum propagation algorithm for lensless holotomography).

|   | **Algorithm S1 – multi-wavelength Gerchberg-Saxton (GS)** |
|---|---|
|   | **Inputs:** $I_{\lambda_1 n}(x,y), I_{\lambda_2 n}(x,y), \alpha_{x_n}, \alpha_{y_n}, \lambda_1, \lambda_2, z_{co}, N_{iter}$ <br> **Output:** $C_{\lambda_1 n}(x,y)$ |
| 1 | $C_{\lambda_1 n}(x,y) = \sqrt{I_{\lambda_1 n}(x,y)}$    % Initial guess |
| 2 | **for:** $n_i = 1: N_{iter}$ |
| 3 | $\quad C_{\lambda_2 n}(x,y) = cross\_wavelength\_transform\{C_{\lambda_1 n}(x,y), z_{co}, \lambda_1, \lambda_2, \alpha_{x_n}, \alpha_{y_n}\}$ |
| 4 | $\quad C_{\lambda_2 n}(x,y) = \frac{C_{\lambda_2 n}(x,y)}{|C_{\lambda_2 n}(x,y)|}\sqrt{I_{\lambda_2 n}(x,y)}$    % Update with collected image |
| 5 | $\quad C_{\lambda_1 n}(x,y) = cross\_wavelength\_transform\{C_{\lambda_2 n}(x,y), z_{co}, \lambda_2, \lambda_1, \alpha_{x_n}, \alpha_{y_n}\}$ |
| 6 | $\quad C_{\lambda_1 n}(x,y) = \frac{C_{\lambda_1 n}(x,y)}{|C_{\lambda_1 n}(x,y)|}\sqrt{I_{\lambda_1 n}(x,y)}$    % Update with collected image |

## Supplementary note 5: Filtered backpropagation LHT reconstruction

Algorithm S2 present the implementation of the filtered backpropagation (FB) algorithm. The algorithm takes as input a set of $N$ complex optical fields $C(x,y)$ acquired for different illumination angles $\alpha_x, \alpha_y$ and a $\lambda_1$ wavelength, along with measured volume axial sampling $\Delta z$ and a user specified range of propagation distances $z_{min}, z_{max}$ that correspond to minimal and maximal axial distance between sample volume and camera. The algorithm relies on propagating each $n^{th}$ optical field: firstly at $-z_{co}$ distance with SS-TR-AS method (mire accurate for longer propagation distances) and then within $z_{min}, z_{max}$ range with SP-TR-AS method (necessary to account for transversal shift for volume reconstruction). For each given $z$, all $N$ propagated fields are averaged to produce the $x,y$ slice of the final $U(x,y,z)$ volume.

|   | **Algorithm S2 – filtered backpropagation (FB) method** |
|---|---|
|   | **Inputs:** $C(x,y), \alpha_x, \alpha_y, \lambda_1, z_{min}, z_{max}, \Delta z$ |

|   | **Outputs:** $U(x, y, z)$ |
|---|---|
| 1 | $U = 0$ |
| 2 | **for:** $n = 1:N$ |
|   | $U_{z_{co}}(x, y) = \text{SS-TR-AS}\{C_n(x, y), -z_{co}, \alpha_{x_n}, \alpha_{y_n}, \lambda_1\}$ |
| 3 |   **for:** $z = z_{min}: \Delta z: z_{max}$ |
| 4 |     $U(x, y, z) = U(x, y, z) + \text{SP-TR-AS}\{U_{z_{co}}(x, y), (z_{co} - z), \alpha_{x_n}, \alpha_{y_n}, \lambda_1\}$ |
| 5 | $U = U/N$ |

## Supplementary note 6: SOLVE algorithm description

Proposed SOLVE algorithm described in detail via pseudo-code presented in Algorithm S3. Firstly, an initial estimate of the 3D complex amplitude distribution is obtained using a modified filtered backpropagation (FB) algorithm (Algorithm S2 in Supplementary Document 1) for the user defined range of propagation distances ($z_{min}$ and $z_{max}$ corresponding to minimal and maximal distance between sample volume and camera) and the measured volume axial sampling $\Delta_z$. The final 3D distribution, $U$, is then refined using the proposed iterative reconstruction algorithm.

Each iteration of the SOLVE algorithm begins with filtering the 3D complex volume $U$ by applying a physical constraint that the normalized object amplitude must be less than or equal to 1 – meaning the object can absorb but not generate light. This step improves the convergence and stability of the reconstruction process. Other constraints could also be applied, such as total variation denoising[29] or assuming that the object's refractive index is either greater or smaller than that of the surrounding medium. Next, for each illumination angle $n$, the multi-plane volume $U$ is propagated to the camera plane using a multi-slice beam propagation method (BPM) model. The propagation between individual slices of U is realized with the use of specialized off-axis shift-preserving AS method (SP-TR-AS), while propagation between last slice of U and camera is realized with shift-suppressed AS (SS-TR-AS). Both methods were originally adopted by us to lensless large volume reconstruction with oblique illumination regime. Alternatively to previous works in this context[22,32] offering general views, we provide scrupulous analysis and rigorous justification for the algorithm steps and their implementation within lensless regime. The difference $\Delta C_n$ between the estimated field at the camera ($C_n'$) and the measured complex amplitude $C_n$ (retrieved from GS) is then calculated. After computing the set of differences $\Delta C_1$ to $\Delta C_N$, a 3D correction volume $\Delta U$ is reconstructed using the FB method. This $\Delta U$ is subtracted from the current $U$ with a learning rate $\varepsilon$, resulting in the updated estimate.

The parameter $\varepsilon$ is selected empirically by observing reconstruction quality over subsequent iterations: too large $\varepsilon$ value leads to oscillations in the background phase, while too small $\varepsilon$ value results in minimal improvement. In this study, $\varepsilon$ was empirically set between 0.005 and 0.025, depending on the dataset, which is generally a recommendable range. The SOLVE algorithm was typically run for five iterations, as no significant improvement was usually achieved within that range.

|    | **Algorithm S3 – SOLVE** |
|----|---|
|    | **Inputs:** $C(x,y), \alpha_x, \alpha_y, \lambda_1, z_{min}, z_{max}, \Delta_z, N_{iter}, \epsilon$ |
|    | **Outputs:** $U(x,y,z)$ |
|    | **% Initial guess (FB)** |
| 1  | $U(x,y,z) = \text{FB}(C(x,y), \alpha_x, \alpha_y, \lambda_1, z_{min}, z_{max}, \Delta_z)$ |
|    | **% Iterative reconstruction** |
| 2  | **for:** $n_i = 1:N_{iter}$ |
|    | **% Absorption constraint** |
| 3  | $A(x,y,z) = |U(x,y,z)|$ |
| 4  | $A(A > 1) = 1$ |
| 5  | $U(x,y,z) = A(x,y,z) \cdot \frac{U(x,y,z)}{|U(x,y,z)|}$ |
|    | **% Forward propagation (multi-slice BPM)** |
| 6  | **for:** $n = 1:N$ |
| 7  | $U_z(x,y) = U(x,y,z_{max})$ |
| 8  | **for:** $z = z_{max}: -\Delta_z: z_{min}$ |
|    | **% Propagate to next plane** |
| 9  | $U_{z-\Delta_z}(x,y) = \text{SP-TR-AS}(U_z(x,y), -\Delta_z, \alpha_{x_n}, \alpha_{y_n}, \lambda_1)$ |
|    | **% Actualize field with next field plane** |
| 10 | $U_{z-\Delta_z}(x,y) = U_{z-\Delta_z}(x,y) \cdot U(x,y,z-\Delta_z)$ |
|    | **% Propagate to camera plane** |
| 11 | $C_n'(x,y) = \text{SS-TR-AS}(U_z(x,y), -z_{min}, \alpha_{x_n}, \alpha_{y_n}, \lambda_1)$ |
|    | **% Calculate difference between estimation and measurement** |
| 12 | $\Delta C_n(x,y) = C_n'(x,y) - C_n(x,y)$ |
|    | **% 4. Backward propagation (FB)** |
| 13 | $\Delta U(x,y,z) = \text{FB}(\Delta C(x,y), \alpha_x, \alpha_y, \lambda_1, z_{min}, z_{max}, \Delta_z)$ |
|    | **% 5. Update U** |
| 14 | $U(x,y,z) = U(x,y,z) - \varepsilon \cdot \Delta U(x,y,z)$ |